\documentclass{egpubl}
\usepackage{eurovis2025}

\EuroVisEducation              

\usepackage[T1]{fontenc}
\usepackage{dfadobe}

\usepackage{cite}  
\BibtexOrBiblatex
\electronicVersion
\PrintedOrElectronic
\ifpdf \usepackage[pdftex]{graphicx} \pdfcompresslevel=9
\else \usepackage[dvips]{graphicx} \fi

\usepackage{egweblnk}

\usepackage{xcolor}

\definecolor{customblue}{HTML}{9ac0e4}
\definecolor{custompink}{HTML}{f48dcb}
\definecolor{customred}{HTML}{e66c6c}
\usepackage{soul}

\title[From Reality to Recognition: Evaluating Visualization Analogies for Novice Chart Comprehension]%
      {From Reality to Recognition: Evaluating Visualization Analogies for Novice Chart Comprehension}

\author[O. Huang \& P.Y.K. Lee \& C. Nobre]
{\parbox{\textwidth}{\centering O. Huang\orcid{0009-0007-1585-1229}, P.Y.K. Lee\orcid{0000-0002-3385-5756}, and C. Nobre\orcid{0000-0002-2892-0509}
        }
        \\
{\parbox{\textwidth}{\centering Department of Computer Science, University of Toronto, Canada
       }
}
}

\begin{document}

\teaser{
 \vspace{-3mm}
 \includegraphics[width=.7\linewidth]{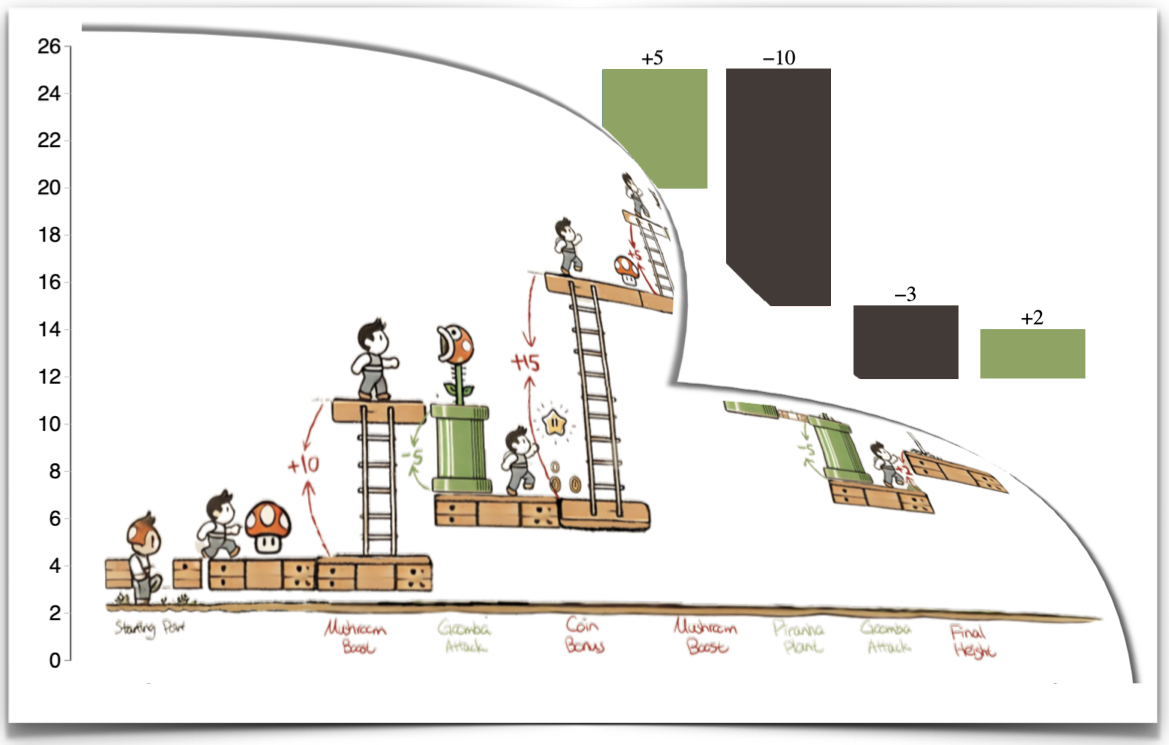}
 \centering
  \caption{Example use of a visualization analogy (top) providing an accessible entry point to understanding the underlying data visualization, a waterfall chart (bottom). The Mario analogy preserves the main features of a waterfall chart (common baseline in adjacent bars) to teach novice viewers how to read the chart. }
\label{fig:teaser}
}

\maketitle
\begin{abstract}
   Novice learners often have difficulty learning new visualization types because they tend to interpret novel visualizations through the mental models of simpler charts they have previously encountered. Traditional visualization teaching methods, which usually rely on directly translating conceptual aspects of data into concrete data visualizations, often fail to attend to the needs of novice learners navigating this tension. To address this, we conducted an empirical exploration of how analogies can be used to help novices with chart comprehension. We introduced visualization analogies: visualizations that map data structures to real-world contexts to facilitate an intuitive understanding of novel chart types. We evaluated this pedagogical technique using a within-subject study (N=128) where we taught 8 chart types using visualization analogies. Our findings show that visualization analogies improve visual analysis skills and help learners transfer their understanding to actual charts. They effectively introduce visual embellishments, cater to diverse learning preferences, and are preferred by novice learners over traditional chart visualizations. This study offers empirical insights and open-source tools to advance visualization education through analogical reasoning.


\end{abstract}  
\section{Introduction}

\vspace{-2mm}
As visualizations increasingly mediate our understanding of everything from scientific discoveries to daily news, the ability to interpret them effectively has become a fundamental skill \cite{nobre2024reading}. However, while most people are familiar with basic charts like bar and line graphs, learning to work with more advanced visualizations remains challenging \cite{bach2023challenges}. A key barrier is that new visualization types often appear arbitrary and disconnected from learners' existing knowledge \cite{ruchikachorn_learning_2015}.

\vspace{-1mm}
One promising but understudied approach to bridging this gap is the use of analogies - mapping new concepts to familiar ones to facilitate learning. While analogies have proven effective in STEM education \cite{harrison2006teaching} and computer science teaching \cite{harsley2016incorporating}, their application to visualization education remains mostly unexplored.

\vspace{-1mm}
In this work, we investigate the effectiveness of visualization analogies: visualizations that map data structures (e.g., histogram bins) to familiar real-world objects (e.g., stacks of exam papers). These analogies aim to make abstract visualization concepts more concrete by grounding them in learners' existing experiences. The aim of our work is to provide an empirical account of whether visualization analogies can be used as an intuitive pedagogical tool in visualization education. Specifically, we conducted a survey with 128 visualization novices \cite{burns_who_2023} on their performance, preference, and learning of one of eight new chart types ranging from the simple (e.g., bar charts) to the complex (e.g., sunburst charts).

\vspace{-1mm}
We focused on \textbf{5 key dimensions} to assess the role of visualization analogies in learning: the \textbf{effectiveness} of visualization analogies on visual analysis tasks; the ability of visualization analogies to support \textbf{learning transfer} to baseline chart visualizations; the influence of \textbf{visual embellishments} in visualization analogies on user attention and potential misinterpretation of charts \cite{bateman_useful_2010, andry_interpreting_2021}; the \textbf{generalizability} of visualization analogies to different learning styles (as captured through participant VARK preferences \cite{aunhabundit_visualizing_2024}); and the self-reported \textbf{preferences} of participants towards visualization analogies.  Our results show that visualization analogies are effective in helping visualization novices in identifying key data encodings while maintaining clarity and intepretability.

\vspace{-1mm}
Overall, the contributions of our work are two-fold:

\textbf{1. Empirical Evaluation of Visualization Analogies.} 
We conducted an empirical study with 128 participants to investigate how mapping chart structures to familiar, real-world objects affects novices’ understanding of data. Our findings show that visualization analogies significantly improve comprehension compared to standard chart representations.

\vspace{-1mm}
\textbf{2. Open Sourcing of Visualization Analogies} 
We develop a set of visualization analogies\cite{analogies}, refining them through feedback from visualization experts. The final set of visualization analogies are made publicly available through an open-source repository (\href{https://visanalogy.github.io/visualization-analogies/}{visanalogy.github.io}), enabling educators and researchers to adapt and extend these techniques.



\vspace{-2mm}
\vspace{-2mm}
\section{Related Work}

\vspace{-2mm}
In this section, we survey two key research areas which inform our approach: efforts to make visualization education more engaging and accessible, and the use of analogies to aid data comprehension.

\vspace{-2mm}
\subsection{Visualization Education for Novices}
Devising pedagogical strategies to help novice learners of visualization techniques adapt to unfamiliar visualization types can be challenging. Chang et al.\cite{chang_strategies_2024} observed that students often chose familiar visualizations for decision-making tasks rather than those best suited to the task. This aligns with Grammel et al.'s findings that novices rely on familiar visualization types when addressing new data visualization needs \cite{grammel2010information}. At the educator level, this means that visualizations that seem intuitive to work with can be highly dependent on the learner's prior experiences. For example, Chevalier et al. \cite{chevalier_observations_nodate} identified a discrepancy among elementary school educators, who see visualizations as intuitive learning tools, yet feel their students are not skilled enough to interpret and create them.

\vspace{-1mm}
Researchers have explored various approaches to overcome this barrier, including: Interactive tools and games (e.g., Roboviz \cite{adar_roboviz_2022}, Diagram Safari \cite{gabler_diagram_2019}), hands-on activities with data physicalization \cite{perin2021students}, and narrative-based approaches \cite{huynh_designing_2020}. These techniques emphasize models of learning that are exploratory, experiential, and curiosity-driven to heighten engagement over a traditional didactic learning model. 

\vspace{-1mm}
While these approaches have shown promise in engaging learners, they often require significant development effort and time investment from learners. Additionally, most focus on a limited range of visualization types, making them difficult to scale across the broad spectrum of charts that novices need to learn.

\vspace{-1mm}
These limitations in existing approaches highlight the need for a teaching method that is both scalable and intuitive. Analogies—a well-established technique in other educational domains—offer a promising direction. By mapping visualization structures to familiar real-world concepts, we can potentially help novices grasp new chart types without the substantial onboarding overhead of games or physical tools. This potential synergy between analogical thinking and visualization education remains largely unexplored, despite its successful application in other learning contexts.
 
 \vspace{-3mm}
\subsection{Analogies in Data Comprehension}

\vspace{-1mm}
Analogies have long been a popular pedagogical technique in other disciplines \cite{harrison2006teaching}, but their use in visualization education has been limited. Studies have demonstrated how abstract numerical data can be made more accessible by re-expressing it through familiar objects and scenarios \cite{chevalier_using_2013, hullman_improving_2018}. While visual analogies \cite{spezzini_effects_nodate} have proven particularly effective for novices by enabling intuitive interpretation of complex data values, these approaches typically focus on single data points rather than complete datasets or visualization structures.

\vspace{-1mm}
More promising developments have emerged in spatial contexts, where researchers have successfully used extended analogies to represent multiple data values. Kim et al. \cite{kim_generating_2016} showed how analogies about area and distance enhance understanding of spatial data. Similarly, Riederer et al. \cite{riederer_put_2018} demonstrated that comparing unfamiliar statistics to familiar object measurements—such as relating country-level statistics to city areas—significantly improved comprehension. Their empirical results confirmed that such perspective-based analogies make complex numerical information more relatable and meaningful. However, scaling these techniques to heterogeneous datasets while maintaining their intuitive nature remains challenging \cite{hullman_improving_2018}.

\vspace{-1mm}
Visual analogies \cite{lin_effectiveness_1996} have further advanced analogical learning by mapping abstract data onto physical forms of everyday objects, making data relationships more apparent \cite{marzano_classroom_2001}. One study found that presenting health data as visual analogies led to better patient understanding compared to conventional line graphs \cite{reading_turchioe_visual_2020}. Another showed that bar chart analogies improved comprehension over traditional representations \cite{morrill_breaking_2017}. While these studies reveal the potential of visual analogies for clarifying complex relationships, they remain limited to simple chart types and lack comprehensive empirical validation for building broader visualization literacy.

\vspace{-1mm}
Recent advances in this space include \textit{AnalogyMate}\cite{chen_beyond_2024}, which systematically links numerical values to familiar objects through a structured design space. Their framework organizes analogy creation strategies into categories like comparison, proportion, unitization, and accumulation—demonstrated through examples such as comparing bar heights to familiar objects. This work extends analogical representations beyond individual measurements, suggesting the potential for whole-dataset comprehension.

\vspace{-1mm}
Our research builds upon these foundations while addressing a specific challenge in visualization education. Rather than focusing on individual numerical measurements, we develop visualization analogies that encompass entire datasets. By mapping visualization structures to real-world objects, we leverage people's existing knowledge to help them understand novel visualizations directly, instead of requiring comprehension through simpler chart types. Our work provides empirical evidence for the effectiveness of this approach in developing visual analysis skills.

\vspace{-3mm}
\section{Visualization Analogies Design}
\subsection{Design Criteria} \label{criteria}
\vspace{-1mm}
To ensure meaningful evaluations, we design visualization analogies that effectively support novice learners in understanding complex concepts. Our design objective is to guide learners to envision concepts in practical situations while preserving the analytical strength of the original charts. Below, we outline the design requirements for the analogies, emphasizing how they contribute to creating intuitive and accessible visualizations.

\vspace{-1mm}
\textbf{Representation of Data Attributes}: Analogies should maintain the data attributes of the original dataset while presenting them in a more accessible and familiar real-world context\cite{chen_beyond_2024}. This helps learners connect abstract chart representations to everyday scenarios, clearly communicating key relationships, trends, and patterns in the data.

\vspace{-1mm}
\textbf{Visual Reconstruction}: Adapting from the concept of re-expression \cite{chevalier_using_2013}, the visual structure of an analogy should closely resemble the original chart. This requires careful attention to layout, shape, and data point representation. The primary aim is to maintain a strong visual correspondence with the original chart. 

\vspace{-1mm}
\textbf{Analysis Potential}: An effective analogy does more than map data into the shape of physical entities. It should maintain meaningful contexts that relate closely to learners' real-world experiences. The chosen scenarios should align with the data's characteristics and maintain conceptual relevance. This connection helps learners link the analogy context to the original data.

\vspace{-1mm}
\textbf{Scalability}: Lastly, visualization analogies must support the same data complexity as the original chart \cite{richer_scalability_2024}. Achieving this requires selecting meaningful and scalable objects as contexts. These choices should naturally accommodate more data points without overwhelming the learner. We chose physical objects that support proportional mapping to achieve this goal. This ensures that larger datasets can be represented intuitively.

\vspace{-3mm}
\subsection{Chart Selection}

\begin{figure}[htb]
  \centering
  \includegraphics[width=1\linewidth]{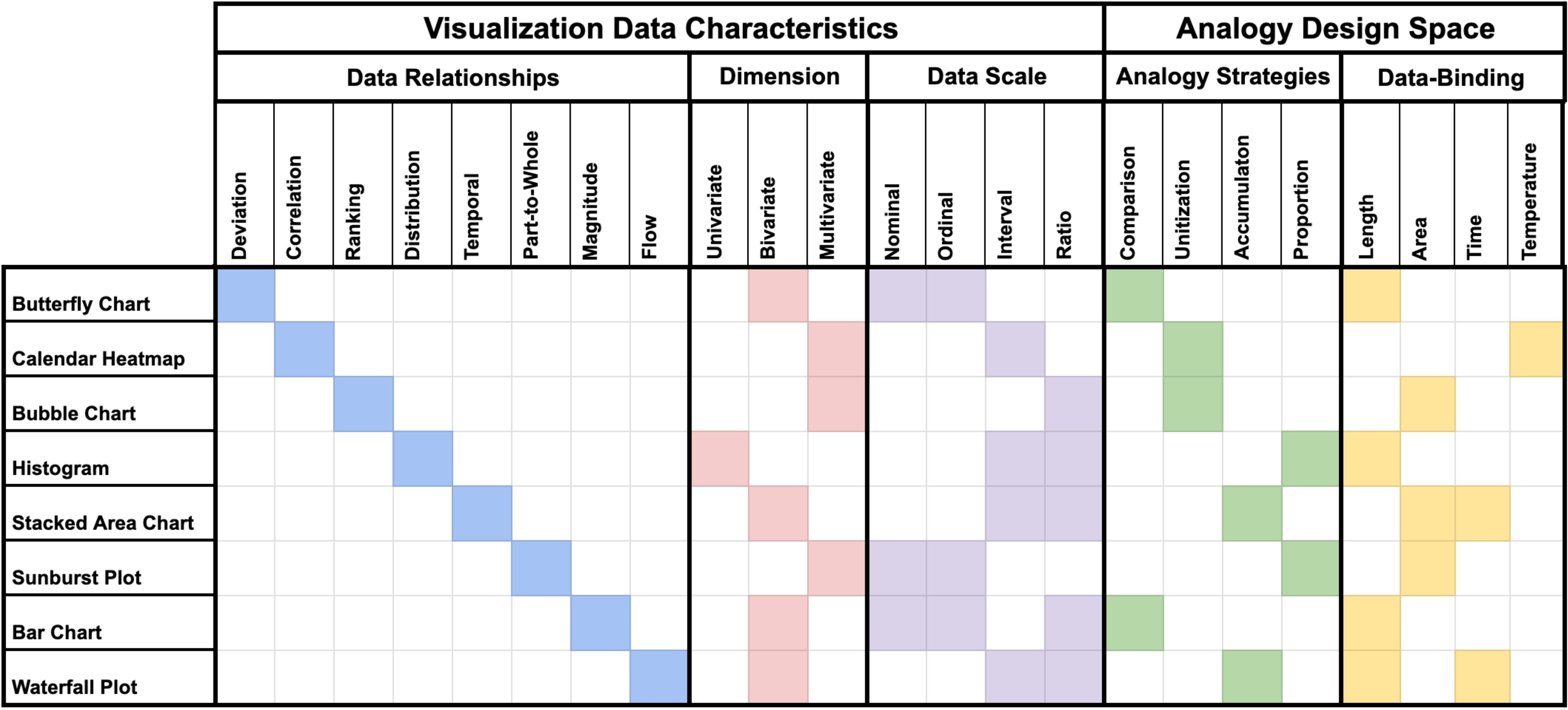}
  %
 
  %
  \caption{\label{fig:selection}
           Chart Selection Framework for Analogy Design}
\end{figure}

\vspace{-1mm}
We selected a diverse set of representative charts from the \href{https://gramener.github.io/visual-vocabulary-vega/}{Visual Vocabulary} to evaluate the effectiveness of analogies (Figure~\ref{fig:selection}). Each chart type represents a unique data relationship, ensuring our analogies can scale across different datasets and analytical needs.

\vspace{-1mm}
Dimensionality affects the complexity of data visualization. To expose learners to various visual complexities, we included charts for univariate, bivariate, and multivariate datasets \cite{denis_spss_2018}, making our analogy framework adaptable to diverse data structures.

\vspace{-1mm}
Data scales also shape data visualization and comprehension \cite{DVLD}. We assessed visualization analogies across scales such as categorical distinctions, ranks, quantitative differences, and proportional relationships. Our chart selection aligns with the established design space for data analogies \cite{chen_beyond_2024}, focusing on comparisons, accumulations, and proportions. We increased chart adaptability using data-binding elements like length or area for quantitative data. Spatial data relationships (e.g. choropleth maps) were excluded as they naturally align with real-world scenarios \cite{borkin_what_2013} and could dilute our unique analogical design approach, despite their commercial popularity \cite{https://doi.org/10.1111/cgf.14031}.

\vspace{-3mm}

\subsection{Visualization Analogy Creation Process}
\vspace{-1mm}
\begin{figure*}[htb]
  \centering

  \includegraphics[width=1\linewidth]{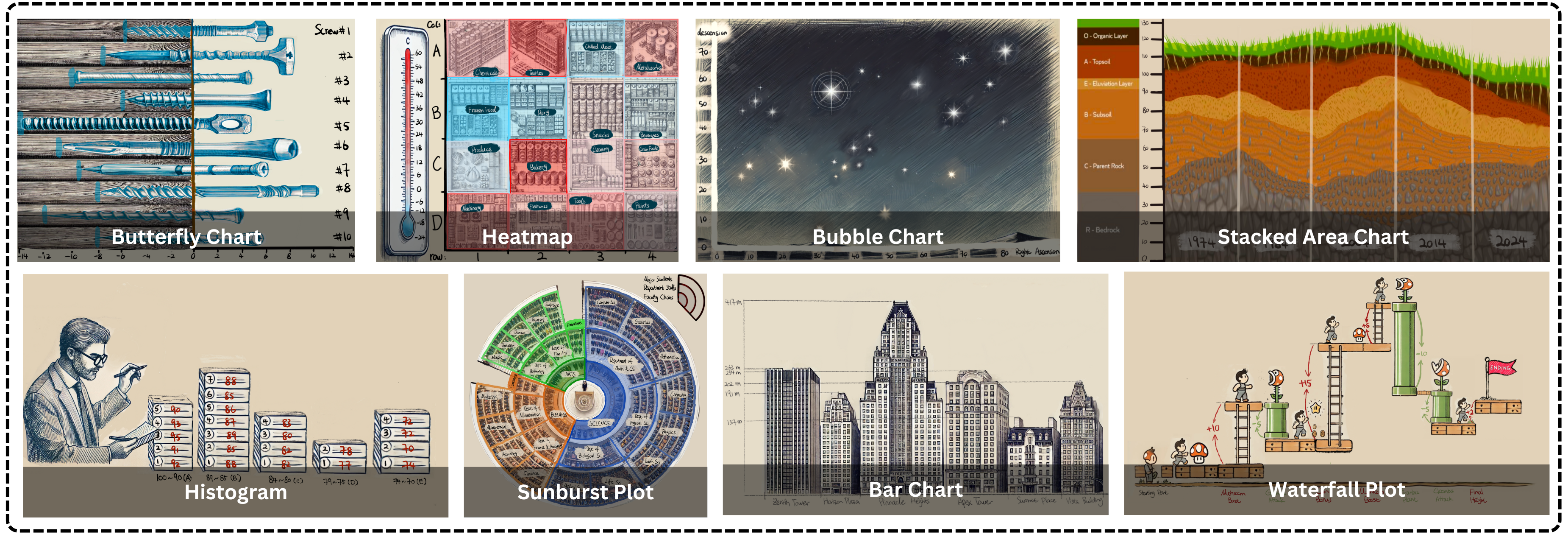}

  \caption{\label{fig:analogies}
           Overview of the 8 selected visualization analogies for the study, chosen based on their alignment with chart characteristics, contextual relevance, and scalability across diverse datasets.}
\end{figure*}

\vspace{-1mm}
\textbf{Challenges in analogy creation:} 
The creation of our visualization analogies (Figure~\ref{fig:analogies}) followed the design criteria outlined in Section \ref{criteria}, with the overarching goal of ensuring clarity, relevance, and scalability across various datasets. Choosing a suitable analogy for each chart type proved challenging. Each proposed analogy had to faithfully capture the chart’s core analytical structure while remaining both conceptually engaging and capable of accommodating larger or more complex datasets. For example, an analogy might excel at explaining quantitative differences but struggle to depict proportional relationships. We adopted an iterative design process to address these trade-offs by systematically evaluating candidate analogies for mapping fidelity (i.e., preserving data attributes), contextual relevance (for novice learners), and scalability (for complex data). This approach ensured that our final designs upheld the design criteria outlined in Section~\ref{criteria}, while still offering an intuitive learning experience.

\vspace{-1mm}
\textbf{Iterative design and expert review:} 
Our iterative process involved initial brainstorming sessions followed by multiple refinement cycles to ensure each visualization analogy met our design requirements. The analogies were then presented to five established visualization researchers, all tenure-track faculty, for formal expert review. These reviewers used a rubric derived from the design criteria \ref{criteria} to evaluate each analogy. Their feedback surfaced issues about the need to preserve proportional relationships, layout clarity, and potential visual clutter when scaled to larger datasets. Insights from the expert reviews were integrated into subsequent design iterations, leading to the final analogies in Figure~\ref{fig:analogies}.

We have open-sourced the final visualization analogies as a novel teaching technique for educational use and future research. They can be found at {\href{https://visanalogy.github.io/visualization-analogies/}{visanalogy.github.io}}

\vspace{-3mm}
\section{Study}
\vspace{-1mm}
Building upon the selected visualization analogies, we conducted a multi-dimensional within-subjects study with 128 visualization novices to evaluate the effectiveness of analogies compared to unmodified charts. The study investigates their potential as an effective teaching technique for understanding and analyzing novel visualizations. In this context, we use the term \textbf{baseline} charts to refer to conventional visualizations without any visual embellishments, serving as a standard for comparison. Our goal was to evaluate the effectiveness of visualization analogies in the following dimensions:

\vspace{-1mm}
\begin{itemize}
    \item \textbf{RQ1 [Performance \& Load]:} How do visualization analogies impact understanding, performance, and cognitive load compared to baseline charts?
    \item \textbf{RQ2 [Learning Transfer]:} Can visualization analogies aid in transferring knowledge on how to interpret similar baseline charts? 
    \item \textbf{RQ3 [Visual Embellishments]:} How do visual embellishments in analogies affect users' attention to data encodings?
    \item \textbf{RQ4 [Generalizability]:} Are visualization analogies effective across different learning preferences?
    \item \textbf{RQ5 [Preferences]:} How do learner preferences affect the effectiveness of visualization analogies in educational contexts?
\end{itemize}

\vspace{-3mm}
\label{subsec:deployment}
\subsection{Study Deployment}

\vspace{-1mm}
The study was conducted on \href{https://www.qualtrics.com}{Qualtrics}. We implemented a think-aloud protocol to gather insights into participants' sentiments, cognitive processes, and decision-making strategies. This allowed participants to record up to two minutes of verbal input for each question. Voice input was captured using the Phonic AI API and integrated into the survey platform.  

\vspace{-1mm}
Additionally, participants completed reproduction tasks by sketching each chart’s key characteristics to demonstrate understanding using a custom drawing interface that offered adjustable pen width and color selection. All study materials —including the Qualtrics questionnaires, sketching platform code, visualization tasks, and evaluation rubric — are publicly available on \href{https://github.com/hivelabuoft/AnalogyVis}{GitHub}, which also contains our expert reviews, deployed visualization analogies, unprocessed think-aloud data, and analysis code for reproducibility. 

\vspace{-3mm}
\subsection{Procedure}
\begin{figure}[htb]
  \centering
  \includegraphics[width=1\linewidth]{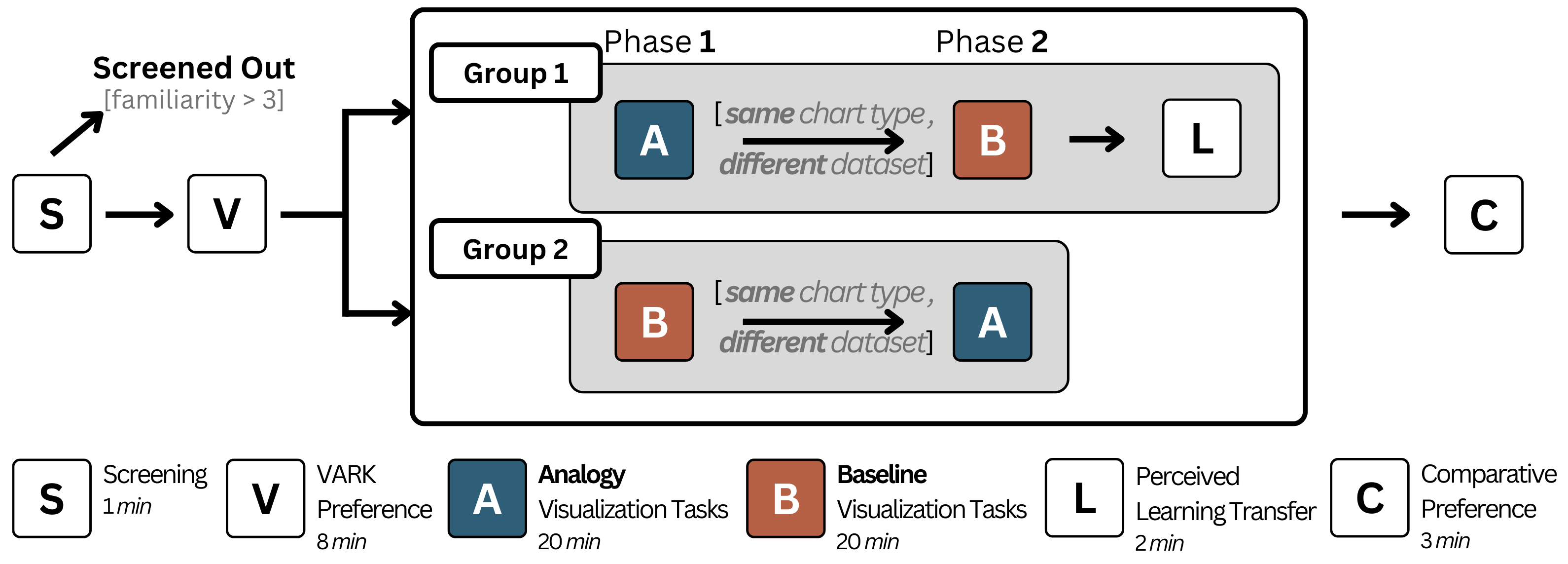}
  \caption{\label{fig:procedure}
           Study procedure overview: participant progression through screening, VARK questionnaire, division into 2 balanced groups, and finally preference questionnaire.}
\end{figure}

\vspace{-1mm}
An overview of the study procedure is displayed in Figure ~\ref{fig:procedure}.

\vspace{-1mm}
The study employed a within-subjects design, with participants evenly distributed into two groups to counterbalance the order of visualization exposure. Group 1 viewed visualization analogies first, followed by baseline charts, while Group 2 experienced the reverse order. The datasets used in each phase were designed to share identical contexts but not the same data, reducing the potential for learning effects from repeated exposure to identical datasets.

\vspace{-1mm}
Initially, each participant was assigned to a specific chart type. They then self-rated their familiarity with the assigned chart on a scale from 1 to 5. To ensure a novice-level comprehension of the given data visualization, only the participants who rated their familiarity at 3 or below were allowed to proceed. To ensure balanced representation and reduce biases associated with specific chart types, the study distributed the selected participants evenly across the 8 chart types. 

\begin{figure}[htb]
  \centering
  \includegraphics[width=1\linewidth]{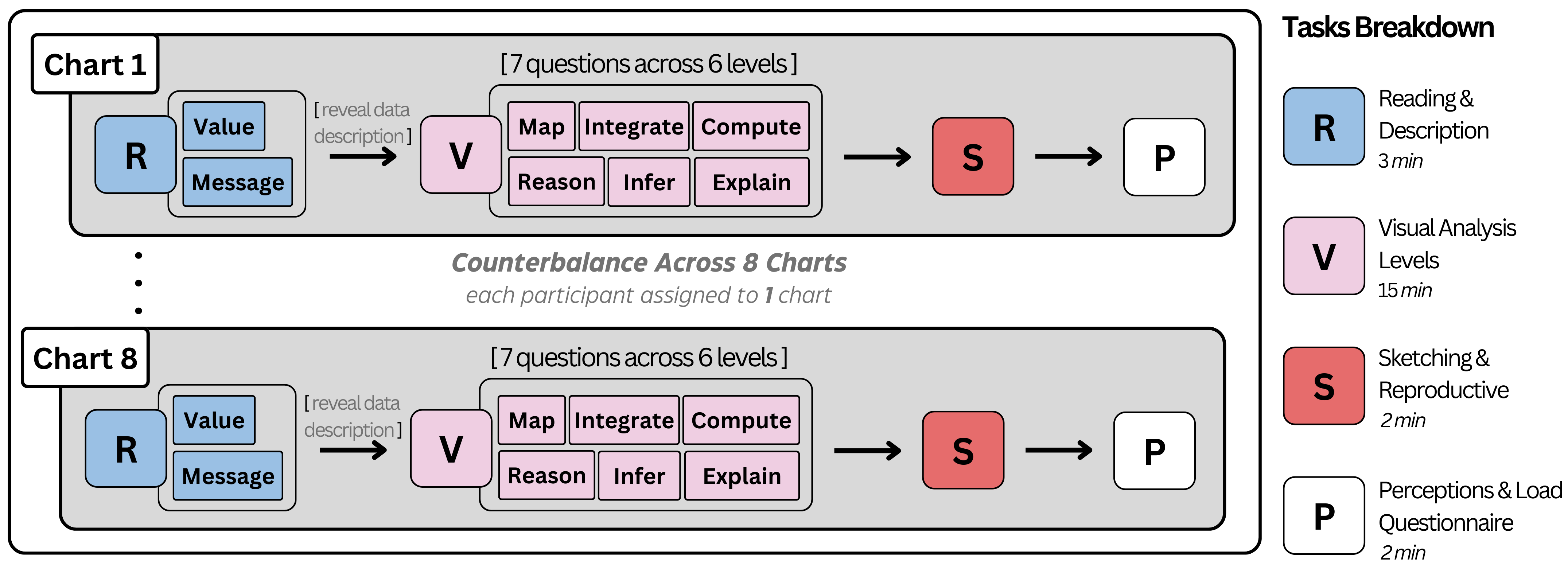}
  \caption{\label{fig:procedure2}
           Breakdown of visualization tasks: reading \& description, visual analysis, sketching, and post-task ratings. }
\end{figure}

\vspace{-1mm}
Figure~\ref{fig:procedure2} illustrates the procedure for each phase. Each phase began with participants viewing a visualization without any accompanying title or description. They were asked to interpret the chart’s purpose and identify key data points based solely on its visual representation (\colorbox{customblue}{Reading \& Description}). These questions were adapted from the framework established by Bateman et al. \cite{bateman_useful_2010} to determine whether visual embellishments might lead to misinterpretation. After that, the chart title and dataset description were provided, and participants proceeded to perform a series of \colorbox{custompink}{Visual Analysis} tasks. These tasks were organized in ascending analytic levels \cite{locoro_visual_2021}, progressing through stages of Mapping, Integrating, Computing, Reasoning, Inferring, and Explaining.

\begin{figure}[htb]
  \centering
  \includegraphics[width=1\linewidth]{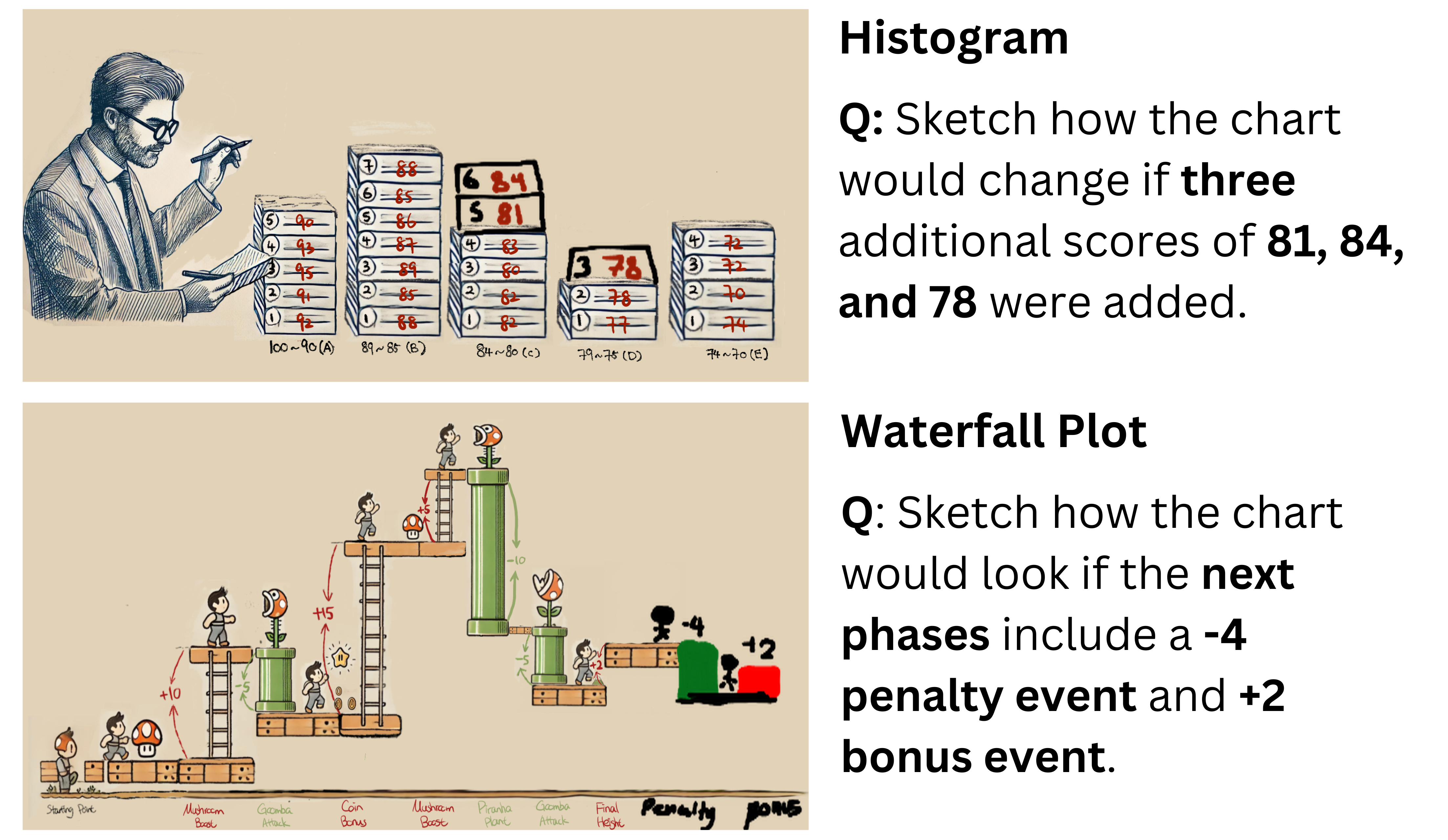}
  \caption{\label{fig:exampleSketch}
           Example sketch questions showcasing visual recreations of a histogram chart (top) and a waterfall plot (bottom)}
\end{figure}

\vspace{-1mm}
Participants were then prompted with a \colorbox{customred}{Sketching} task (Figure~\ref{fig:exampleSketch}) where we aim to evaluate their comprehension and retention of the presented data. Participants ended the survey with a post-task questionnaire to evaluate their perceived learning and confidence levels. Additionally, the NASA-TLX was employed in the Perceptions \& Load section to assess the cognitive load - explicitly focusing on mental demand, effort, and frustration encountered while interpreting each visualization.

\vspace{-1mm}
For participants in Group 1 (i.e, for those who saw the analogy first and then the baseline),  we included additional questions to assess their self-rated learning transfer to the baseline chart across interpretation, context, and data relationships. Finally, all participants compared the two visualization representations, evaluating them in terms of intuitiveness, engagement, and clarity.

\vspace{-3mm}
\subsection{Participants}

\vspace{-1mm}
After screening for minimal familiarity with the tested chart types, we recruited 128 participants through \href{https://www.prolific.co/}{Prolific} (79 males and 49 females), aged 19–68 (M = 31.4). The final sample had an average familiarity rating of 2.09 out of 5, reflecting a novice-level understanding of data visualization. They completed the VARK questionnaire to determine their learning preferences, focusing on their dominant modality\cite{VARKGuide}: Visual (n = 30), Auditory (n = 28), Reading/Writing (n = 31), and Kinesthetic (n = 39). The study received approval from our institute's ethics review board, and informed consent was obtained from all participants before the study. We paid participants \$15 to participate in the study. On average, each participant took 1 hour to complete the study. 

\vspace{-3mm}
\vspace{-2mm}
\subsection{Data Analysis}

\vspace{-1mm}
Each visualization task was given equal weight. We assessed participants' responses using a detailed rubric focused on the completeness and justification of their reasoning. We conducted an intra-rater reliability test, where the first author re-evaluated a randomly selected subset of responses after a 3-week interval. To confirm the consistency of the rubric, we measured the agreement between the initial and re-evaluated scores using Cohen’s Kappa. Each question was scored out of 5, resulting in a maximum possible score of 40 points for the visualization task, and 10 points for description tasks. We also recorded the time spent on each question to analyze cognitive engagement and compare the efficiency of visualization analogies to baseline charts.


\vspace{-1mm}
For performance scores, we employed a mixed-effects model to control for fixed effects, such as visualization technique, order of exposure, chart difficulty, and random effects associated with individual participant differences. For statistical analysis, we employed a one-way ANOVA for normally distributed data with equal variances and applied the  Kruskal-Wallis H test to non-normal data. We conducted pairwise comparisons using Welch's t-tests for normally distributed data and Mann-Whitney U tests for non-normal data.


\vspace{-3mm}
\section{Results}
\vspace{-1mm}
In this section we report on the effectiveness of visualization analogies across the 5 measured dimensions. First, we evaluate how these analogies impact performance in visual analysis tasks compared to baselines \textbf{[RQ1]}. Subsequently, we investigate their role as a teaching tool by exploring the learning transfer effects from analogy charts to baselines \textbf{[RQ2]}. We also examine how visual embellishments in analogies influence users' attention to data encodings \textbf{[RQ3]}. Furthermore, we assess the generalizability of visualization analogies across VARK learning preferences \textbf{[RQ4]}. Finally, we capture learners' perceptions and analyze their engagement and comprehension preferences regarding these visualization methods \textbf{[RQ5]}.

\begin{figure*}[htb]
  \centering
  \includegraphics[width=1\linewidth]{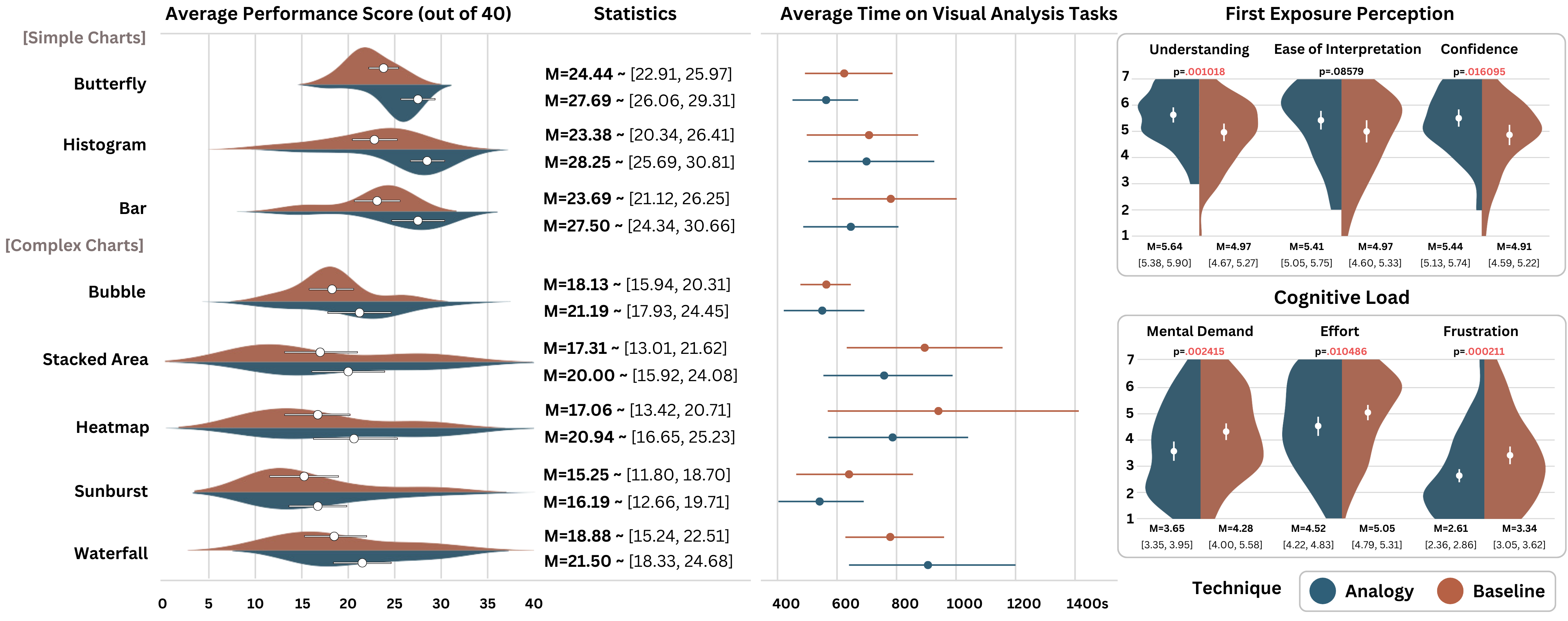}
  \caption{\label{fig:rq1result}
           An overview of visual analysis tasks results (from left to right): the average performance score for each technique; the average time spent on each technique; self-reported evaluations on perceptions of understanding, ease of interpretation, and confidence from initial exposure; and cognitive load evaluations of mental demand, effort, and frustration based on the NASA-TLX assessment.}
\end{figure*}

\vspace{-2mm}
\subsection{RQ1 [Performance \& Load]: Effectiveness of Visualization Analogies}

\vspace{-1mm}
We investigated the average performance score across all visualization tasks (Figure~\ref{fig:rq1result}). The result indicates \textbf{visualization analogies significantly enhance users' performance in visual analysis tasks compared to baselines.} Participants who used visualization analogies achieved higher performance scores (\textit{M=22.91, SD=7.30}) than those who used baselines (\textit{M=19.77, SD=6.65}). A paired t-test revealed a statistically significant difference (\textit{p=.0004}). 

\vspace{-1mm}
The selection of charts used in this study include charts across a spectrum of complexities. In order to investigate the effect of analogies as a function of complexity, we performed a cluster analysis on the performance scores to categorize the charts based on difficulty levels. The optimal clustering resulted in two groups, allowing us to separate the charts into \textbf{simple} and \textbf{complex} categories.

\vspace{-1mm}
Building on these categorizations, we found that \textbf{the effectiveness of analogies can vary based on chart complexity.} For simple charts, participants using analogy charts had a superior mean performance score of \textit{27.81 (SD=4.66)}, compared to \textit{23.83 (SD=4.54)} for baselines. This difference was highly significant (\textit{p<.0001}). For complex charts, the analogy condition yielded a mean score of \textit{19.96 (SD=7.03)}, while the baseline condition had a mean of \textit{17.33 (SD=6.54)}. While the difference in performance between the two conditions was not as high for complex charts, it was still a significant difference (\textit{p = .015}).

\vspace{-1mm}
Our mixed-effects model further confirmed these findings. The fixed factors included visualization technique, chart difficulty, average time spent, and order of exposure. The model showed that analogy charts led to an average of \textit{11\%} higher performance scores compared to baselines. Simpler charts outperformed complex ones by \textit{19\%}. Notably, the average time spent on tasks was not a significant factor, indicating that participants could understand analogy charts quickly and still demonstrate a high level of comprehension.

\vspace{-1mm}
In regards to the self-reported dimensions, \textbf{visualization analogies led to significantly higher comprehension and confidence compared to baselines.} Participants felt they understood the charts better when first exposed to analogy charts. Comprehension ratings were significantly higher \textit{(p=0.001)}. Confidence ratings were also significantly higher \textit{(p=0.016)} for analogy charts \textit{(M=5.41)} than baselines \textit{(M=4.97)}. Ratings for ease of interpretation did not reach statistical significance \textit{(p=0.086)}, suggesting that both methods maintain a certain level of cognitive friction necessary for extracting insights. 

\vspace{-1mm}
To further explore the overall load introduced by analogies, we examined cognitive load across three dimensions: mental demand, effort, and frustration. The result indicates \textbf{visualization analogies reduce cognitive load across all three measured dimensions.} Specifically, participants experienced lower mental demand when using analogy charts \textit{(p=0.002)}. They reported exerting less effort \textit{(p=0.010)} and feeling less frustration \textit{(p<0.001)} compared to baseline charts. These measures suggest that visualization analogies improve performance and make analysis tasks less cognitively demanding.


\vspace{-2mm}
\subsection{RQ2 [Learning Transfer]: Enhancing Comprehension from Analogy to Baseline}

\begin{figure}[htb]
  \centering
  \includegraphics[width=1\linewidth]{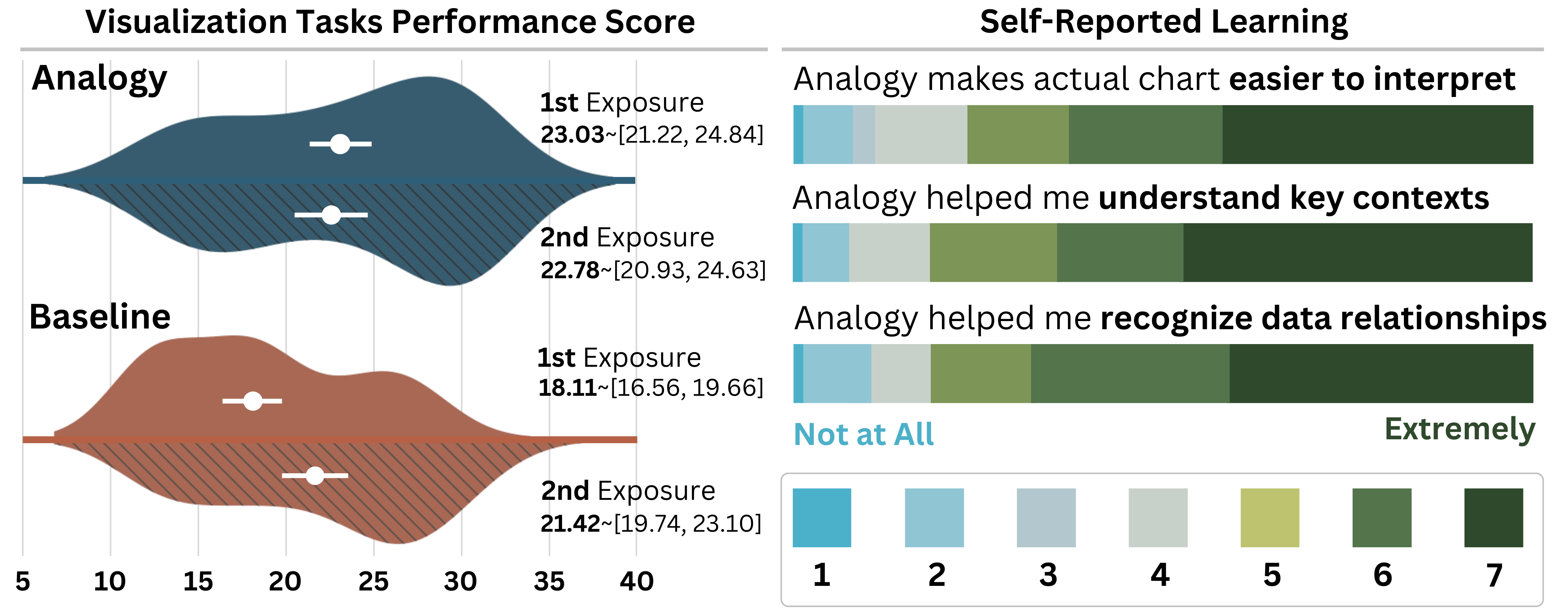}
  \caption{\label{fig:rq3}
           Overview of learning transfer result. (left) performance score by first and second exposure to each technique; (right) self-rated perceived learning transfer on a scale of 1 to 7}
\end{figure}

\vspace{-1mm}
In order to establish the ability of analogies to promote learning transfer, we first examined the performance scores for analogy charts across both groups with varying exposure orders, finding that \textbf{analogies are effective independently, regardless of prior exposure to baselines.} Participants performed similarly on analogy charts, regardless of whether or not they had previously seen the baseline charts. As shown in Figure~\ref{fig:rq3}, participants who were first exposed to analogy charts had a mean performance score of \textit{23.03 (SD=7.22)}, while those who viewed analogy charts after the baselines had a mean score of \textit{22.78 (SD=7.44)}. This difference was insignificant \textit{(p=0.929)}. 

\vspace{-1mm}
Given that analogy charts are not affected by potential learning effects from the baseline charts, we examined how prior exposure to analogies influenced performance to understand the impact on learning transfer. Our result indicates \textbf{visualization analogies promote learning transfer to baseline charts.} Participants who saw analogy charts before baselines performed significantly better on the baselines. The mean performance score for baselines after analogy exposure was \textit{21.42 (SD=6.73)}, compared to \textit{18.11 (SD=6.19)} for those who saw baselines first. This improvement was statistically significant \textit{(p=0.007)}. This indicates that analogy charts enhance understanding and aid in interpreting baseline charts.

\vspace{-1mm}
Aligning with the performance score, \textbf{participants reported high levels of perceived learning transfer from analogy to baselines.} A majority of participants felt that analogy charts helped them interpret baselines more efficiently. Specifically, 76.56\% reported that analogies made it easier to interpret the baseline. Additionally, 81.25\% said analogies helped them understand key contexts of the baseline, and 81.25\% felt analogies helped them recognize data relationships more clearly.

\vspace{-1em}

\subsection{RQ3 [User Attention]: Understanding the Effects of Visual Embellishment}

\vspace{-1mm}
In order to investigate whether visual embellishments (an intrinsic part of analogies) detracted from viewer attention and comprehension of the charts, we investigated users' ability to interpret the assigned chart before being given any information on the underlying dataset. We call this metric `interpretation accuracy'. Our findings indicate that \textbf{visualization analogies do not compromise users' attention to data encoding}. Participants' descriptions of the analogy charts were as accurate as those of the baseline charts. Even without the chart title or dataset context, they could identify the core idea using both visualization techniques. The mean description scores for the analogy \textit{(M=9.01)} and baseline \textit{(M=8.36)} were similar, and a pairwise t-test showed no significant difference between the techniques \textit{(p=0.872)}.

\vspace{-1mm}
Further analysis examined the impact of chart complexity on interpretation accuracy. \textbf{For complex charts, visualization analogies significantly enhanced interpretation accuracy compared to baselines}. Participants scored higher using visualization analogies \textit{(M=8.51, SD=3.05)} than with baselines \textit{(M=7.16, SD=2.95)} when describing charts with more challenging interpretations. This statistically significant difference \textit{(p=0.047)} suggests that the additional visual elements provided by the analogies facilitated better interpretation of complex charts. In contrast, for simpler charts, the average scores were comparable between analogy \textit{(M=9.56, SD=1.79)} and baseline \textit{(M=9.56, SD=1.79)}. The pairwise t-test showed no significant difference \textit{(p=0.087)}, indicating that both techniques were equally effective in conveying key information in simpler visualizations.

\subsection{RQ4 [Generalizability]: Effectiveness Across Diverse Learning Preferences} \label{results:generalizability}

\begin{figure}[htb]
  \centering
  \includegraphics[width=1\linewidth]{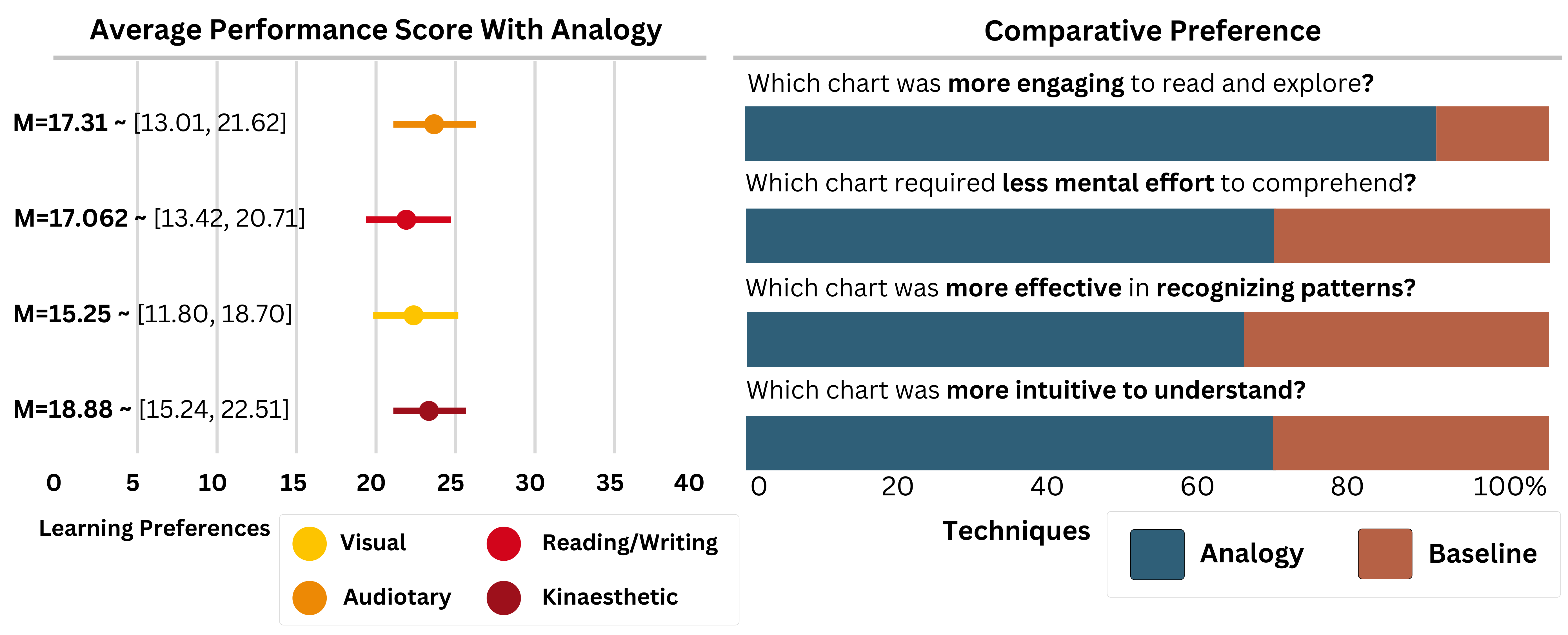}
  \caption{\label{fig:rq5}
           (left) Performance score with analogies across all learning preferences. (right) Comparative preference results: each bar displays the percentage of participants favoring analogies versus baseline charts for each of the displayed questions.}
\end{figure}

\vspace{-1mm}
We conducted an ANOVA F-test to compare the performance scores grouped by different learning preferences. We found that \textbf{visualization analogies are generalizable teaching techniques, effective across various learning preferences.} There was no significant difference in performance scores among the VARK learning preference categories when using analogy charts \textit{(p=0.826)}. Mean scores were similar (Figure~\ref{fig:rq5}): Auditory (\textit{22.63, SD=7.27}), Kinesthetic (\textit{21.87, SD=7.08}), Reading/Writing (\textit{20.16, SD=7.06}), and Visual (\textit{20.72, SD=7.12}). These results indicate that visualization analogies are equally effective regardless of individual learning preferences.

\vspace{2mm}

\vspace{-4mm}
\subsection{RQ5 [Preference]: Engagement and Comprehension Preferences for Visualization Analogies} 

\vspace{-1mm}
We examined participants' perceptions of visualization analogies in terms of engagement and found that \textbf{visualization analogies are perceived as more engaging.} In terms of engagement, \textit{83.59\%} of participants found visualization analogies more engaging than standard charts, with one noting that \textit{"it feels like playing a game"}. This playful quality appears to resonate across diverse learning preferences. In regards to comprehension, \textit{70.31\%} reported that analogies required less mental effort to interpret, citing clearer visuals that streamline data processing and reduce cognitive load. Additionally, \textit{63/28\%} of  participants also highlighted that analogies aided in pattern recognition, further indicating their potential to clarify complex data through intuitive design elements.

\vspace{-2mm}
\section{Discussion}
\vspace{-1mm}
Our study demonstrates that visualization analogies can effectively enhance novice performance in visual analysis tasks while facilitating learning transfer to baseline charts. We highlight four main points for discussion. 

\vspace{-2mm}
\subsubsection*{Generalizability Across Learning Styles}

\vspace{-1mm}
 Visualization analogies maintained their effectiveness across different VARK learning preferences, as evidenced by consistent performance scores regardless of self-reported learning styles. This suggests visualization analogies can have broad appeal in classroom environments, supporting diverse learners without privileging any particular learning approach. This universality is especially valuable in visualization education, where students often possess vastly different prior experiences and levels of familiarity with visualizations. Our results indicate that visualization analogies can leverage what appears to be the inherently intuitive nature of visualizations as teaching tools \cite{chevalier_observations_nodate} while actually translating this into developed visual literacy skills across the learning spectrum. We also speculate that analogies can be applied differently depending on learning style to target a learner's particular preferences and bolster the efficacy of visualization analogies. For example, kinesthetic learners may prefer visualization analogies that explicitly incorporate an analogy to real-world movement (e.g., Mario analogy for a waterfall chart), so they can more comfortably map from their preferred learning style to an abstract concept.


\vspace{-3mm}
\subsubsection*{Strengths and Limitations in Skill Development}

\vspace{-1mm}
Visualization analogies proved most effective when used for descriptive or diagnostic reasoning tasks, making them particularly valuable for developing fundamental chart reading and interpretation skills. For instance, participants excelled at chart reading tasks (e.g., \textit{"How many storage sections have a temperature above 30°C?"} in the heatmap) but struggled with questions requiring extrapolation beyond the visible analogy (e.g., \textit{"If aisle C's temperature increased to match aisle A, how might this change the overall temperature distribution in the storage area and what items may no longer be suitable in aisle C?"}). This suggests that while analogies excel at building foundational understanding, additional development may be needed to support more complex analytical skills. One promising direction is the potential use of animated visualization analogies to help learners develop extrapolation abilities with datasets.

\vspace{-3mm}
\subsubsection*{The Double-Edged Nature of Visual Embellishments}

\vspace{-1mm}
While our visualization analogies generally avoided introducing unnecessary chart junk \cite{bateman_useful_2010}, we found that the effectiveness of visual embellishments varies significantly with learners' prior experiences. Depending on a learner's familiarity with the analogy context, there is a risk that visual elements intended to support the analogy might be misinterpreted as meaningful data channels. For example, in our Mario Game visualization analogy of the waterfall plot, some participants interpreted the chart as being about \textit{"how to get a better score in Mario,"} confusing the contextual elements with the visualization's actual purpose.

\vspace{-1mm}
Even subtle design choices can generate confusion. In our histogram analogy, the decision to stack exam booklets in descending order—despite histograms having no inherent ordering requirement—led some participants to incorrectly conclude that the visualization represented \textit{"the sorted ranking of marks for students [...] organized from the students who scored the highest marks to the lowest marks."} These observations highlight how visual embellishments can significantly impact interpretation based on viewers' prior experiences. Success requires carefully balancing the elements needed to create meaningful analogical connections while avoiding superfluous details that might confuse novice learners.

\vspace{-2mm}
\subsubsection*{Challenges with Complex Visualizations}

\vspace{-1mm}
We observed particular challenges when applying analogies to multivariate charts like sunburst diagrams. While our goal was to help novices understand novel visualizations without referencing simpler chart types, participants often reverted to interpreting complex visualizations in terms of more familiar ones when faced with completely unfamiliar structures. For instance, some participants described the sunburst chart as a \textit{"pie chart [that] has colours to better represent the sections visually"} likely due to visual similarities with familiar pie charts. Similarly, participants often described histograms in terms of bar charts, noting that the histogram \textit{"is a bar graph with no labeled axis, but from left to right, the bars are starting from down changing to upwards."}

\vspace{-1mm}
Despite these challenges, visualization analogies demonstrated clear benefits, particularly when chart titles and labels were omitted. Participants showed better immediate comprehension with analogies compared to baseline charts. For example, one participant could describe how a stacked area chart analogy \textit{"probably [showed] how much of this material increase or decrease over the years and how the ground is later shaped"} while being unable to interpret the baseline version \textit{("I can't tell")}. This suggests that even when observers are completely unfamiliar with a visualization type, analogies can provide sufficient contextual information to extract meaningful information without external assistance.

\vspace{-1mm}
These findings indicate that while visualization analogies offer significant promise for visualization education, their successful implementation requires careful attention to: (1) balancing analogical elements with data clarity; (2) supporting the progression from basic comprehension to complex analysis; (3) selecting appropriate and accessible real-world contexts; (4) managing visual complexity based on learner expertise; and (5) providing additional support for complex multivariate visualizations.

\vspace{-1mm}
Our work provides empirical evidence for the effectiveness of visualization analogies while highlighting important considerations for their practical implementation in visualization education.


\vspace{-2mm}
\section{Conclusion}
\vspace{-1mm}
This paper introduced visualization analogies as a novel teaching technique that maps datasets to real-life objects, sharing visual and conceptual characteristics with the data. We present an initial exploration of their effectiveness as tools for enhancing comprehension in visual analysis tasks. The goal of making abstract information more intuitive and relatable is to improve user understanding when learning new chart types.

\vspace{-1mm}
Our empirical investigation revealed that visualization analogies significantly improve user performance, comprehension, and confidence while being more engaging and less cognitively demanding than traditional charts. The visual embellishments in the analogies were particularly beneficial for navigating complex charts, did not mislead users, and were generalizable across diverse learning styles. However, learners still require support when using visualization analogies for complex multivariate charts, such as sunburst charts. Future work should explore their long-term impact on retention and explore self-supervision and friction-reduction methods to enhance understanding of complex analogies. We hope this work will inspire further research into the development and application of visualization analogies, enhancing their role as standalone and complementary tools in educational contexts.

\clearpage




\section*{Acknowledgments}

This work was supported by the Natural Sciences and Engineering Research Council of Canada (NSERC). We would also like to thank Emily Wall, Jillian Aurisano, Yalong Yang, Alark Joshi, and Lane Harrison for their invaluable feedback and expert evaluation throughout the course of this research.

\bibliographystyle{eg-alpha-doi} 

\newcommand{\etalchar}[1]{$^{#1}$}


\newpage


\end{document}